\documentclass[prb, twocolumn, superscriptaddress,showpacs, floatfix]{revtex4}
\usepackage{epsfig}
\usepackage{amssymb, amsmath, bbm}
\usepackage{color}

\newcommand{\beq}{\begin{equation}}
\newcommand{\eeq}{\end{equation}}
\newcommand{\beqa}{\begin{eqnarray}}
\newcommand{\eeqa}{\end{eqnarray}}
\newcommand{\bec}{\begin{center}}
\newcommand{\eec}{\end{center}}

\newcommand{\bs}{\boldsymbol}

\begin{document}
\title{Frustrated spin ladder with alternating spin-1 
and spin-1/2 rungs}
\author{V. Ravi Chandra}
\affiliation{Max-Planck-Institute for Physics of Complex Systems,
N\"{o}thnitzer Str-38, D-01187, Dresden, Germany}
\affiliation{Physics Department, The Technion, Haifa, 32000, Israel}
\author{N. B.  Ivanov}
\affiliation{Fakult\"at f\"ur Physik, Universit\"at Bielefeld,
D-33501 Bielefeld, Germany}
\affiliation{Institute of Solid State Physics,  Bulgarian Academy of
Sciences, Tzarigradsko chaussee 72, 1784 Sofia, Bulgaria}
\author{J. Richter}
\affiliation{Institut f\"ur Theoretische Physik, Universit\"at Magdeburg,
PF 4120, D-39016 Magdeburg, Germany}

\date{\today}
\begin{abstract}
We study the impact of the diagonal frustrating couplings on the quantum
phase diagram of a two-leg ladder composed of
alternating spin-1 and spin-1/2 rungs. As the coupling strength is increased the
system successively exhibits two gapped paramagnetic phases 
(a rung-singlet and a Haldane-like non-degenerate states) and
two ferrimagnetic phases with different ferromagnetic moments per rung.
The first two states  are similar to the phases studied in the frustrated 
spin-1/2 ladder, whereas the magnetic phases appear as a result of  the
mixed-spin structure of the model. A detailed characterization of these
phases is presented using density-matrix
renormalization-group calculations, exact diagonalizations of
periodic clusters, and an effective Hamiltonian approach inspired by 
the analysis of numerical data. The present theoretical study was motivated
by the recent synthesis of the  quasi-one-dimensional ferrimagnetic material
Fe$^{II}$Fe$^{III}$ (trans-1,4-cyclohexanedicarboxylate) exhibiting a
similar ladder structure.
\end{abstract}
\maketitle
\section{Introduction}
Over the past two decades there has been an increasing interest  
in  quantum  spin systems  with competing  exchange
interactions.\cite{diep,scholl} 
Quantum  spin chains and ladders with frustration, both 
for half-integer and integer spins, set up an important part of this 
research since they provide a unique testing ground based on the available
powerful 
analytical and numerical techniques for one-dimensional (1D) systems.
In particular, the frustrated ladder models have allowed controlled 
calculations to examine topological order,\cite{white} dimer 
order,\cite{starykh}  as well as the appearance of
fractional  excitations in spin models.\cite{allen}
Most of previously studied frustrated chain and ladder models have been 
related to uniform-spin structures with all the spins same. 
In comparison, till now much less experimental as well as  theoretical 
work concerning the impact of competing interactions in quasi-1D  mixed-spin 
systems has been accomplished.\cite{ivanov_09} Often these  systems exhibit 
quasi-1D ferrimagnetic ground states  with 
a net ferromagnetic moment, so that apart from rich quantum phase diagrams  
they might be expected to provide generic examples of 1D magnetic-paramagnetic
quantum phase transitions.\cite{sengupta} 

\begin{figure}[hbt]
\samepage
\begin{center}
\includegraphics[width=7.5cm]{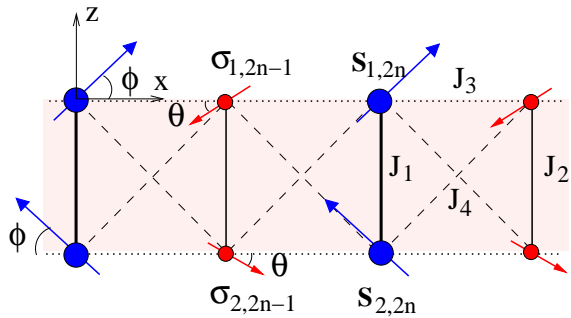}
\vspace{-.4cm}
\caption{\label{ladder_f1} The mixed-spin ladder considered in the paper.
The arrows show the classical canted state described by the angles
$0<\phi<\pi/2$ and $0<\theta <\pi/2$ for the classical spins with magnitudes 
$S_1$ and $S_2$, respectively. The other two classical phases correspond
to spin configurations with $(\phi,\theta)=(0,0)$ 
(antiferromagnetic state) and $(\phi,\theta)=(\pi/2,\pi/2)$
(ferrimagnetic state). 
}
\vspace{-.7cm}
\end{center}
\end{figure}

On the experimental side, during the  past two decades it  has become  
possible to  synthesize a large variety of  quasi-1D materials with 
ferrimagnetic properties. Most of  these materials are 
heterometallic  molecular magnets containing different transition metal 
ions in the unit cell.\cite{kahn} A generic spin model describing these materials
is the quantum Heisenberg spin chain with  antiferromagnetic nearest-neighbor 
exchange interactions  and  two types of alternating quantum spins with 
magnitudes $S_1$ and $S_2$ ($S_1>S_2$).\cite{pati,brehmer,ivanov_98} 
In the extreme quantum case  of spins $(1,1/2)$, the latter model was shown 
to provide an excellent description of the  thermodynamic parameters  of 
the  recently synthesized quasi-1D  bimetallic  compound 
NiCu(pba)(D$_2$O)$_3\cdot$2D$_2$O 
(pba = 1,3-propylenebis).\cite{hagiwara1}
Another important class of quasi-1D ferrimagnets -- the so-called topological 
ferrimagnets -- is related to some homometallic materials exhibiting  composite
chain  structures with different magnetic sublattices.\cite{nishide} 
The homometallic  material A$_3$Cu$_3$(PO$_4$)$_4$ (A=Ca,Sr,Pb)
is an example of  such quasi-1D  ferrimagnets: In this compound, the Cu$^{2+}$ 
ions form   diamond chains  with strongly coupled trimers bridged by oxygen 
ions.\cite{anderson}  Since quasi-1D homometallic 
materials usually have rich  exchange pathway structures, 
they may be expected 
to provide some real examples of quasi-1D ferrimagnets
with magnetic frustration. To the best of our knowledge, the recently
synthesized mixed-valent magnetic material  Fe$^{II}$Fe$^{III}$
(trans-1,4-cyclohexanedicarboxylate)\cite{zheng} provides
the first real example of a quasi-1D Heisenberg ferrimagnet with magnetic
frustration.\cite{zheng} The experimentally established magnetic
structure for temperatures larger than 36 K corresponds 
to the mixed-spin ladder with diagonal exchange bonds shown in 
Fig.~\ref{ladder_f1}, where the site  spins 
$S_1=5/2$ and $S_2=2$ are respectively related to the magnetic ions 
Fe$^{III}$ and  Fe$^{II}$.\cite{zheng} 

The mentioned experimental achievements motivated a series of theoretical 
studies on quantum mixed-spin chains and ladders with geometric 
frustration.  The symmetric diamond chain
with antiferromagnetic vertical bonds was probably the first studied  
model of a 1D quantum ferrimagnet with competing 
interactions.\cite{takano1} A variant of this  
model, the distorted spin-1/2 diamond chain, has received special
theoretical\cite{okamoto1} as well as experimental\cite{ohta} interest 
due to its rich quantum phase diagram~\cite{tonegawa} 
and the relevance for the real 
material Cu$_3$(CO$_3$)$_2$(OH)$_2$. The diamond Heisenberg chain is also 
one of the simplest quantum spin models admitting four-spin cyclic exchange 
interactions.\cite{ivanov1} A generic quantum spin model of a frustrated
1D ferrimagnet is the mixed-spin Heisenberg chain composed of two types of 
alternating spins interacting via  competing nearest-neighbor 
and  next-nearest-neighbor  antiferromagnetic
exchange bonds.\cite{ivanov2} This model may also be  considered 
as a mixed-spin zigzag ladder and is a ferrimagnetic analogue of  
the frustrated Heisenberg chain with ferromagnetic 
nearest-neighbor and antiferromagnetic  next-nearest-neighbor 
exchange bonds. The  spin-1/2 frustrated $J_1-J_2$ ferromagnetic
chain  has recently attracted much attention,\cite{ferro} as it is supposed to describe 
a number of quasi-1D edge-sharing cuprates, such as 
Rb$_2$Cu$_2$Mo$_3$O$_{12}$\cite{hase}, 
Li$_2$ZrCuO$_4$,\cite{drechsler} and LiCuVO$_4$.\cite{enderle}
The latter material exhibits  multiferroic properties\cite{naito} as well 
as an interesting  specific  phase transition in a magnetic field from an 
ordered  spiral  to an ordered  modulated-collinear  magnetic 
phases.\cite{banks} There are other two generic types of frustrated
mixed-spin ladder models describing two interacting mixed-spin alternating
chains. The first one is the checkerboard mixed-spin Heisenberg ladder with 
frustrating diagonal exchange couplings,\cite{ivanov3} and the second one 
is the two-leg ladder model with two types of alternating rungs presented 
in Fig.~\ref{ladder_f1}. Finally, there has been a lot of recent work
reporting interesting quantum phase diagrams in different composite
Heisenberg chains with ferrimagnetic ground states~\cite{yoshikawa} 

In this study we focus on the effects of frustration on the ground state 
phase diagram of the mixed-spin ladder shown in Fig.~\ref{ladder_f1}. In
addition to the theoretically interesting question of the effects of 
frustration in this system, an experimental realization of a closely 
related system in a mixed-valence iron polymer further motivates us.\cite{zheng} 
In the next section we introduce the model and study some relevant 
properties of its Hamiltonian. In Section~\ref{phases},  
we give a detailed description of the quantum phases by using 
an effective Hamiltonian approach 
inspired by the analysis of data obtained using density-matrix 
renormalization-group (DMRG) and exact diagonalization (ED) techniques. 
We conclude in Section~\ref{sum} with a brief summary 
of the results.
\section{The model}\label{model} 
The system under consideration (see Fig.~\ref{ladder_f1}) consists of two 
equivalent  mixed-spin Heisenberg chains (characterized by the
nearest-neighbor exchange constant $J_3>0$)  coupled via
rung ($J_1,J_2>0$) as well as diagonal ($J_4\geq 0$) exchange
bonds. The Hamiltonian of the system reads as 
\begin{equation}\label{h1}
{\cal H}={\cal H}_{12}+{\cal H}_{3}+{\cal H}_{4},
\end{equation}
where
\begin{eqnarray}
{\cal H}_{12}&=&\sum_{n=1}^{L/2}
\left( J_1\bs{s}_{1,2n}\cdot \bs{s}_{2,2n}+
J_2\bs{\sigma}_{1,2n-1}\cdot\bs{\sigma}_{2,2n-1}\right),\nonumber\\
{\cal H}_{3}&=&J_3\sum_{n=1}^{L/2}\sum_{m=1}^{2}
 \left[\bs{s}_{m,2n}\cdot\left(\bs{\sigma}_{m,2n-1}+\bs{\sigma}_{m,2n+1}\right)\right],\nonumber\\
{\cal H}_{4}&=&J_4\sum_{n=1}^{L/2}
 \left[\bs{s}_{1,2n}\cdot\left(\bs{\sigma}_{2,2n-1}+\bs{\sigma}_{2,2n+1}\right)\right.\nonumber\\
&+&
\left.\bs{s}_{2,2n}\cdot\left(\bs{\sigma}_{1,2n-1}+\bs{\sigma}_{1,2n+1}\right)\right].\nonumber
\end{eqnarray}
Here $\bs{s}_{k,2n}$ and $\bs{\sigma}_{k,2n-1}$ ($k=1,2$) are, respectively,
spin-$S_1$ and spin-$S_2$  operators ($S_1>S_2$), and $L$ is the number of rungs.  

 It is instructive to present the Hamiltonian in the following form
\begin{equation}\label{h}
{\cal H}={\cal H}_{12}+ \sum_{n=1}^{L/2}\left[
J_s\, \bs{s}_{2n}\cdot\left(\bs{\sigma}_{2n-1}
+\bs{\sigma}_{2n+1}\right)\right]
+J_aV\, ,
\end{equation}
where $J_{s,a}= (J_3\pm J_4)/2$, and $\bs{s}_{2n} =\bs{s}_{1,2n}+\bs{s}_{2,2n}$ and
$\bs{\sigma}_{2n+1} =\bs{\sigma}_{1,2n+1}+\bs{\sigma}_{2,2n+1}$ are
rung spin operators.  
The operator  $V$ reads as
\begin{equation}\label{v}
V=\sum_{n=1}^{L/2}
=\bs{L}_{2n}\cdot \left( \bs{l}_{2n-1}+\bs{l}_{2n+1}\right),
\end{equation}
where $\bs{L}_{2n} =\bs{s}_{1,2n}-\bs{s}_{2,2n}$ and
$\bs{l}_{2n\pm 1} =\bs{\sigma}_{1,2n \pm 1}-\bs{\sigma}_{2,2n \pm 1}$ 
are rung vector operators. The following  analysis of the 
zero-temperature quantum phase diagram addresses the extreme 
quantum case  of spins $S_1=1$ and $S_2=1/2$, and is mainly restricted
to the parameter  subspace defined by $J_1=J_2=J_3>0$ and $J_4\geq 0$.
To some extent, such a choice of the parameters is motivated 
by the experimentally established strengths of the exchange couplings 
in the ferrimagnetic ladder material  
Fe$^{II}$Fe$^{III}$ (trans-1,4-cyclohexanedicarboxylate).\cite{zheng} 
\subsection{Symmetries of the model}
The mixed-spin system  inherits some important symmetries of
the  parent uniform-spin Heisenberg ladder with
diagonal interactions.\cite{weihong} First, if  the 
parameters $J_3$ and $J_4$ in ${\cal H}$ are exchanged, one can 
recover the original  Hamiltonian by exchanging either the spins on the $S_1$ rungs
($\bs{s}_{1,2n} \longleftrightarrow \bs{s}_{2,2n}$), or the spins on the
$S_2$ rungs ($\bs{\sigma}_{1,2n-1}\longleftrightarrow
\bs{\sigma}_{2,2n-1}$). This means that 
${\cal H}(J_1,J_2,J_3,J_4)={\cal H}(J_1,J_2,J_4,J_3)$. Therefore, the study
of the model can be restricted in the region $J_4/J_3\leq 1$ since the model
with $J_4/J_3>1$ maps onto the one with $J_4/J_3<1$. Because of
the same symmetry,  the Hamiltonian (\ref{h}) does not contain mixed products
of rung spins and  rung vector operators.

The second property of ${\cal H}$ concerns the subspace $J_3=J_4$
($J_a=0$), when the last term in  Eq.~(\ref{h}) disappears.
As is the uniform-spin case,\cite{gelfand} in this parameter subspace the Hamiltonian 
${\cal H}$  commutes with the  local operators $\bs{s}_{2n}^2$ and $\bs{\sigma}_{2n-1}^2$
($n=1,2,\ldots,L/2$), which means that the rung spins $s_{2n}$ and 
$\sigma_{2n-1}$ [defined as  $\bs{s}_{2n}^2=s_{2n}(s_{2n}+1)$ and 
$\bs{\sigma}_{2n-1}^2=\sigma_{2n-1}(\sigma_{2n-1}+1)$] are good local 
quantum numbers. Thus in  every sector of the Hilbert space, 
  defined by the sequence 
$[\sigma_1,s_2,\ldots,\sigma_{L-1},s_{L}]$, the first two 
terms in Eq.~(\ref{h}) reduce to the constant 
\begin{eqnarray}\label{e0}
\hspace{-1cm}E_0&=&-\frac{L}{2}\left[ J_1S_1(S_1+1)+ J_2S_2(S_2+1)\right]
\nonumber\\
&+&\frac{1}{2}\sum_{n=1}^{L/2} 
\left[J_1s_{2n}(s_{2n}+1)
+J_2\sigma_{2n-1}(\sigma_{2n-1}+1)\right].\nonumber
\end{eqnarray}
Thus  Eq.~(\ref{h}) takes the simple form of a Heisenberg spin chain
\begin{equation}\label{h0}
{\cal H}_0=E_0+\sum_{n=1}^{L/2}
J_s\, \bs{s}_{2n}\cdot\left(\bs{\sigma}_{2n-1}
+\bs{\sigma}_{2n+1}\right) .
\end{equation}
The above expression for $E_0$ implies that for strong enough rung 
interactions ($J_1/J_3,J_2/J_3\gg 1$) the  singlet eigenstate 
of Eq.~(\ref{h0}), defined as a product of local rung-singlet states, 
becomes an exact ground state of the model. This state belongs to
the sector $[0,0,\ldots,0,0]$ and can be considered as a prototype of the
rung-singlet phase of Eq.~(\ref{h}) discussed below. 
The following analysis of the quantum phase diagram of Eq.~(\ref{h}) 
implies that in the extreme quantum limit $(S_1,S_2)=(1,1/2)$
the sectors $[1,1,\ldots,1,1]$, $[1,2,\ldots,1,2]$, and 
$[1,1,1,2,\ldots,1,1,1,2]$ also play
an important role: In the first sector, the model defined by
 Eq.~(\ref{h0}) is equivalent to the spin-1 Haldane chain, whereas in the 
last two sectors  Eq.~(\ref{h0}) represents spin-alternating ferrimagnetic 
chains. The ground states related to these  models appear in the quantum
phase diagram of the discussed system.        
\subsection{Classical phase diagram}\label{GS_class}
The classical phases of Eq.~(\ref{h1}) can be described by the angles $\phi$
and $\theta$ (see Fig.~\ref{ladder_f1}) which determine the orientations of
the classical spins in the $xz$ plane.  We consider the parameter subspace
defined by $J_1=J_2=J_3=1$ and $J_4 \ge 0$.  The expression for the ground-state 
energy per cell containing two rungs is seen to be
\begin{eqnarray}
\frac{E_c}{S_1S_2}&=&-\frac{S_1}{S_2}\cos (2\phi )-\frac{S_2}{S_1}\cos (2\theta ) 
\nonumber \\
&&-4\cos (\phi-\theta ) +4J_4\cos (\phi+\theta ).
\end{eqnarray}
A minimization using the independent angle variables $\phi$ and $\theta$
gives the following equations:
\beqa\label{cl}
\cos (\phi+\theta )&=&\frac{c_1}{\kappa}J_4-c_2\nonumber \\
\cos (\phi-\theta )&=&c_2J_4-c_1\kappa, 
\eeqa
where $c_1=\sigma-\sigma^{-1}$, $c_2=\sigma+\sigma^{-1}$, and
$\sigma = S_1/S_2>1$. The parameter $\kappa =\kappa (J_4)$ reads
$\kappa = (4J_4^2/3-1/3)^{1/2}$. 

The lower ($J^{(d)}_4$) and the upper ($J^{(u)}_4$) phase boundaries 
of the classical canted phase shown in Fig.~\ref{ladder_f1}
are related to the inequalities $|\cos (\phi+\theta )|,|\cos (\phi-\theta )|
\leq 1$  implying
\beqa\label{boundary}
J^{(d)}_4&=&\frac{c_2+1}{\sqrt{4(c_2+1)^2-3c_1^2}}\,,\nonumber\\
J^{(u)}_4&=&\frac{c_2-1}{\sqrt{4(c_2-1)^2-3c_1^2}}\, .
\eeqa
\begin{figure}[hbt]
\samepage
\begin{center}
\includegraphics[width=4cm]{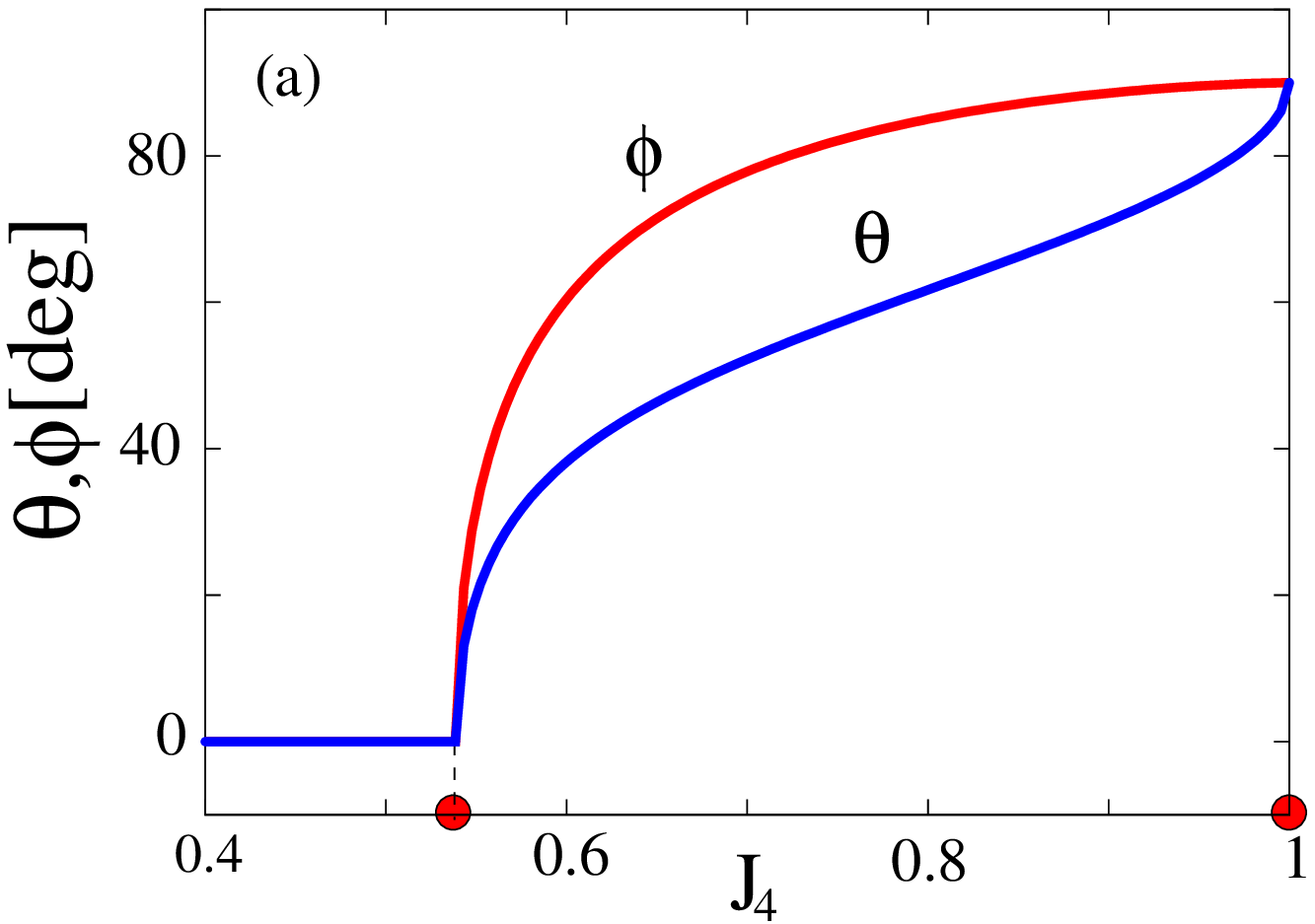}
\includegraphics[width=4cm]{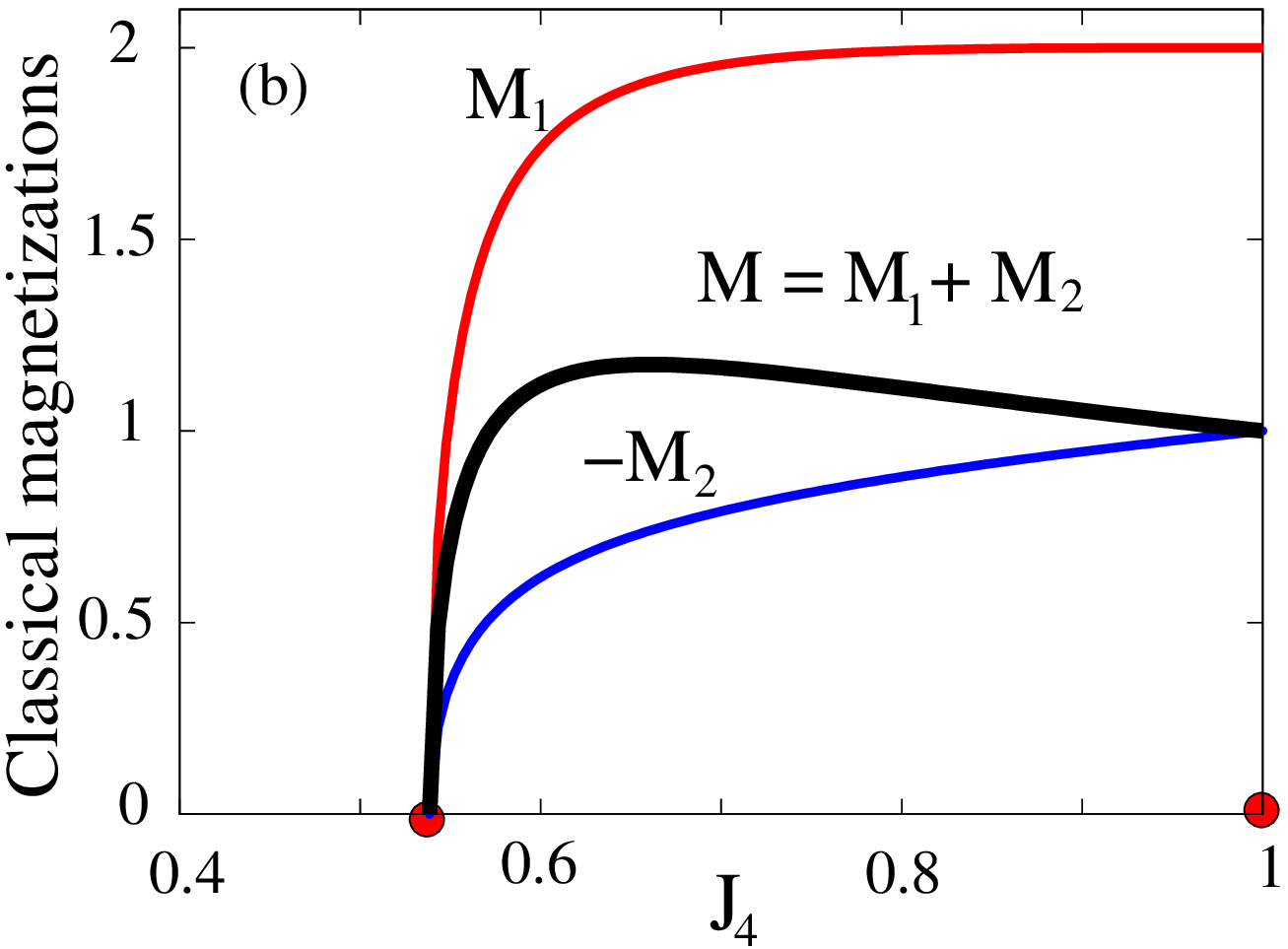}
\vspace{-.4cm}
\caption{\label{classic_f2} (a) The classical phase diagram described
by the angles $\phi$ and $\theta$ \textit{vs.} $J_4$, as obtained from 
Eq.~(\ref{cl}) for the system with $S_1=1$ and
$S_2=1/2$. (b) z components of the classical  magnetizations in the $S_1$
($M_1$) and $S_2$ ($M_2$) sites of the same system. The filled circles on
the $J_4$ axis correspond to the classical transition points
$J^{(d)}_4=7/13$ and $J^{(u)}_4=1$.  
}
\vspace{-.7cm}
\end{center}
\end{figure}                                      

For $J_4<J^{(d)}_4$, we get states of zero magnetization in which the two spins on any rung
and spins along a leg are antiferromagnetically aligned. The
canted state realized for $J^{(d)}_4 < J_4<J^{(u)}_4$ has a
net magnetization that takes a maximal value at some intermediate $J_4$
between both boundaries [see Fig.~\ref{classic_f2}(b)].   
For  $J_4>J^{(u)}_4$ this classical canted phase gives way to a ferrimagnetic state where
all the spins of the same magnitude are ferromagnetically aligned but the relative alignment
of $S_1$ and $S_2$ is antiferromagnetic. Notice that the magnetic
measurements in Ref.~\onlinecite{zheng} indicate  the discussed  ferrimagnetic
configuration--eventually with a small canting of the classical spins-- as the most 
probable spin configuration realized in the real material
Fe$^{II}$Fe$^{III}$. For $S_1=1$ and $S_2=1/2$, 
the above equations give $J^{(d)}_4=7/13\approx 0.538$ and
$J^{(u)}_4=1$. For the  real material studied in
Ref.~\onlinecite{zheng} ($S_1=5/2$, $S_2=2$), one has
$J^{(d)}_4=61/121\approx 0.504$ and
$J^{(u)}_4=21/39\approx 0.553$.     

Interestingly, the discussed classical ferrimagnetic state appears  only
for relatively small values of $\sigma$. For larger $\sigma$,  the lowest
energy collinear configuration for  large $J_4$ is a non-magnetic state
with ferromagnetically arranged legs pointing in opposite directions
(i.e., antiferromagnetically aligned rungs). Comparing the energies of
both configurations ($E_c^{(1)}=S_1^2+S_2^2-4S_1S_2-4S_1S_2J_4$,
$E_c^{(2)}=-S_1^2-S_2^2+4S_1S_2-4S_1S_2J_4$, respectively),
we see that the ferrimagnetic configuration is realized only in the interval 
$1<\sigma \leq 2+\sqrt{3}\approx 3.73$. In the large $\sigma$ case, the
canted phase is also modified: On increasing the parameter  $J_4$ from
$J_4^{(d)}$  up to $J_4^{(u)}$, the $S_2$ spins smoothly 
change their orientation  by  $\pi$, whereas the net orientation
of the larger $S_1$ spins coincides at the phase boundaries. 
In both variants of the classical phase diagram the phase boundaries are defined by 
Eq.~(\ref{boundary}).       

Finally, the  discussed  classical phase diagrams were  independently confirmed
by our classical Monte-Carlo simulations. Below we  argue that 
the classical ferrimagnetic phase survives quantum fluctuations, whereas both 
the antiferromagnetic as well as the canted classical phases are completely
destroyed.\\
\begin{figure}[t,b]
\begin{center}
\includegraphics[width=8cm]{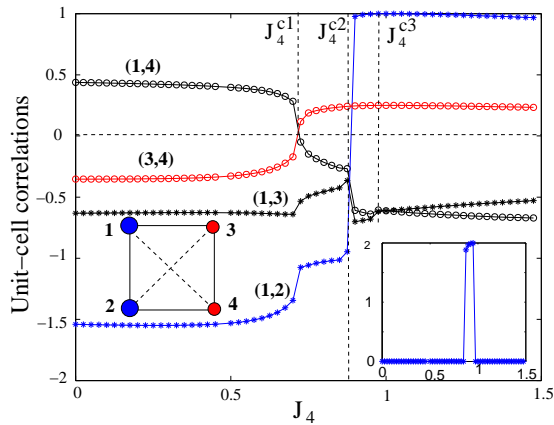}
\caption{\label{c1} Unit-cell isotropic spin-spin  correlations
as a function of the frustration parameter $J_4$, as obtained from the 
DMRG method for open boundary conditions ($L=100$). 
$J_4^{c1}=0.710$, $J_4^{c2}=0.875$, and $J_4^{c3}=0.975$ are the special
points identified as phase-transition points between different ground states. 
The inset shows the  difference in the spin-1 rung correlations in 
two neighboring cells.  Note that the presented
spin-spin correlations belong to unit cells far from the ends.
}
\end{center}
\end{figure}
\section{Quantum phase diagram}\label{phases}
We consider the parameter subspace defined by $J_1=J_2=J_3\equiv 1$ and 
$0\leq J_4\leq 1.5$,
and use the DMRG method\cite{white4} for open boundary conditions 
supplemented by ED data for periodic 
clusters  containing up to $L=14$ rungs. DMRG is carried out for this system
for a range of lattice sizes up to $L=100$ rungs with the spin values $S_1=1$ 
and $S_2=1/2$, respectively. Up to  $320$ density matrix eigenvectors were
retained. Depending on the value of $J_4$, the truncation errors are between
$10^{-7}$ and $10^{-12}$. 

\begin{figure}[t]
\begin{center}
\epsfig{figure=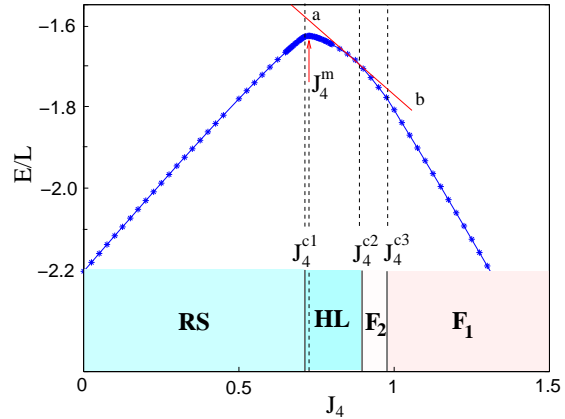,width=8.0cm}
\caption{\label{gse} Ground-state
energy per rung as a function of the frustration parameter $J_4$
(DMRG, $L=90$). $J_4^{m}=0.723$ denotes the location of the maximum.
The positions  of the special points identified in
Fig.~\ref{c1} separate different ground states:
RS (rung-singlet), HL (Haldane-like ), and two different ferrimagnetic
states ($F_1$ and $F_2$). The straight line $ab$ represents the  energy of the
 Haldane state ($|\Psi_H\rangle$) defined by  $E_H=\langle \Psi_H|{\cal
H}|\Psi_H\rangle$. 
}
\end{center}
\end{figure}

The DMRG results presented in Fig.~\ref{c1}  reveal three
special  points on the $J_4$ axis separating  regions with  different 
characteristics  of the short-range correlations: $J^{c1}_4= 0.710$,  
$J^{c2}_4= 0.875$, and $J^{c3}_4= 0.975$.  The same points are
also presented in Fig.~\ref{gse} which shows  DMRG results ($L=90$) for 
the ground-state energy of the mixed-spin model (\ref{h1}). A detailed numerical
analysis,  using both the DMRG  and  
ED methods, predicts  singlet 
ground states in the entire  region $0\leq J_4< J^{c2}_4$.  
For $J_4>J^{c2}_4$, the same analysis suggests ground states  
characterized by net ferromagnetic moments.  Below we argue that these
special  points are related to  quantum phase transitions 
between different ground states.
\subsection{Mapping onto the frustrated spin-1/2 ladder}
An  inspection of the  short-range correlations presented in
Fig.~\ref{c1} implies that the weight of  the local rung quintet (i.e., $s_{2n}=2$)
states on the spin-1 rungs  is negligible almost in the whole interval
$0\leq J_4< J^{c2}_4$. Indeed,  by using the identity 
$\langle \bs{s}_{1,2n}\cdot
\bs{s}_{2,2n}\rangle=\left(\langle  \bs{s}_{2n}^2\rangle
-3/2\right)/2-5/4$ , one finds that the following  relation between the
average rung correlations should be satisfied for any state with a zero  
 weight of the rung quintet states:
 \begin{equation}\label{rcor}
\langle \bs{s}_{1,2n}\cdot
\bs{s}_{2,2n}\rangle =\langle \bs{\sigma}_{1,2n-1}\cdot
\bs{\sigma}_{2,2n-1}\rangle-\frac{5}{4}.   
\end{equation}
As  seen from the numerical results, the above relation is almost perfectly
fulfilled in the entire region $0\leq J_4< J^{c2}_4$, excluding some narrow
vicinity of the point  $J_4^{c2}$ where the correlations 
$\langle \bs{s}_{1,2n}\cdot\bs{s}_{2,2n}\rangle$ abruptly change to $\approx 1$. 
The extremely small contribution of the quintet rung states  in the region
 $0\leq J_4< J^{c2}_4$ can be explained by the peculiarities of the energy
spectrum of the mixed-spin plaquette, where  the lowest 
quintet state  happens to be well separated from the
low-lying triplet and singlet states. Note that the excitation of local
quintet states is  controlled by the last term ($V$) in the
Hamiltonian (\ref{h}). Thus, starting from an eigenstate
belonging to the sector $s_{2n},\sigma_{2n-1} =0,1$ ($n=1,\ldots,L/2$), 
the first-order corrections to the wave function of this eigenstate
will contain relatively small amount of configurations belonging to the 
sectors with local quintet states
due to the larger energy denominator in the perturbation expression.  

 These observations  suggest, in particular,  that in the discussed region the 
ground-state properties  of the mixed-spin system may be approximately
interpreted  by projecting out the local quintet states in the  mixed-spin 
Hamiltonian (\ref{h}). Up to first order in $J_a$, the projected Hamiltonian reads 
as (see the Appendix)
\begin{eqnarray}\label{he}
{\cal H}_{eff}&=&-\frac{5}{8}JL+\sum_{n=1}^{L}\left[ 
J_{\perp}^{'}\bs{\sigma}_{1,n}\cdot \bs{\sigma}_{2,n}\right.\nonumber\\ 
&+& \left. J_s^{'}\, \bs{\sigma}_{n}\cdot\bs{\sigma}_{n+1} 
+\, J_a^{'} \, \bs{l}_{n}\cdot\bs{l}_{n+1}\right],
\end{eqnarray}
where $\bs{\sigma}_{1,n}$ and $\bs{\sigma}_{2,n}$ are spin-1/2 operators, 
$\bs{\sigma}_{n}=\bs{\sigma}_{1,n}+\bs{\sigma}_{2,n}$,
$\bs{l}_{n}=\bs{\sigma}_{1,n}-\bs{\sigma}_{2,n}$, $J_{\perp}^{'}=J_{\perp}$,
$J_s^{'}=J_s$, and $J_a^{'}=-2\sqrt{2/3}\, J_a$. For simplicity, 
we have restricted ourselves to the case of equal rung couplings
($J_1=J_2\equiv J_{\perp}$).
The effective Hamiltonian (\ref{he}) describes a  frustrated spin-1/2
Heisenberg ladder characterized by three parameters, i.e., the strength
of the rung  ($J_{\perp}^{'}$),  leg ($J_3^{'}=J_s^{'}+J_a^{'}$), 
and  diagonal ($J_4^{'}=J_s^{'}-J_a^{'}$) exchange bonds. 
Using the same reasoning, it may be safely suggested that the next-order 
corrections in $J_a$ do not change substantially  the singlet ground states, 
so that the  effective Hamiltonian (\ref{he}) may be used (i) to identify the 
singlet ground states of the original Hamiltonian (\ref{h}) in the region 
$0\leq J_4<J^{c2}_4$ and (ii) to analyze the related quantum phase
transitions.  

As is well-known, as a function of the frustration parameter $J_4^{'}$ 
the model  (\ref{he}) exhibits the so-called rung-singlet (RS) and
Haldane-like (HL) phases.\cite{starykh,weihong,wang,vekua,hung,kim,liu} 
Both ground states are non-degenerate and exhibit finite  singlet-triplet gaps. 
The character of the quantum  RS-HL transition in the weak-coupling limit 
is still under debate: Some of the  cited works\cite{weihong,wang,hung,kim}  suggest  a direct 
first-order transition  between these  phases, but the  others  predict an intermediate columnar 
dimer phase.\cite{starykh,vekua,liu}  Thus the  mapping of  Eq.~(\ref{h}) implies that  
the special point $J_4=J_4^{c1}$ can presumably  be identified 
as a quantum phase transition point separating similar phases. 
Of course, such an  analysis does not exclude the presence  of some intermediate 
singlet phases  in a tiny interval between the RS and HL states. 
Some hints in this direction inspired by the DMRG results for the 
ground-state energy (Fig.~\ref{gse})  will be discussed below in more
detail.

\begin{figure}[hbt]
\samepage
\begin{center}
\includegraphics[width=8cm]{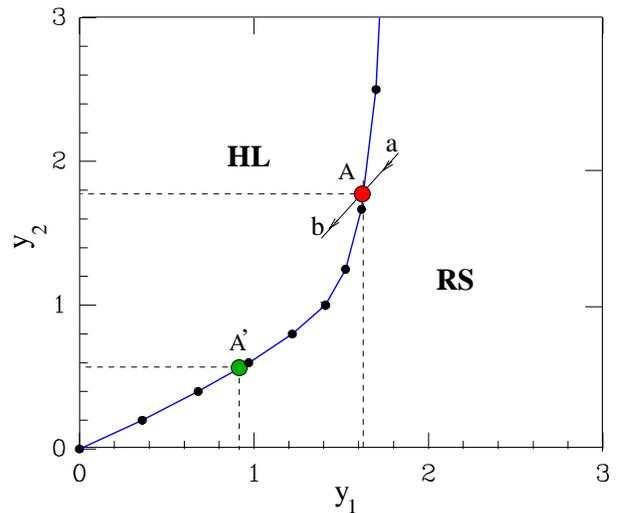}
\vspace{-.4cm}
\caption{\label{map} Phase diagram of the effective spin-1/2 ladder model
with diagonal bonds.\cite{weihong} The  point $A$  with coordinates 
$(y_1,y_2)=(1.618,1.766)$ is the image  of the special point 
$J_4^{c1}=0.710$ obtained by using Eq.~(\ref{b}). The point $A^{'}$ is an image of
 $A$ corresponding  to the symmetry transformation  $J_3^{'}\longleftrightarrow J_4^{'}$.
$ab$ is the path in the $(y_1,y_2)$ plane corresponding to the change
of $J_4$ at fixed $J_{\perp}=J_3=1$.    
}
\vspace{-.7cm}
\end{center}
\end{figure}

The established connection  with  the frustrated spin-1/2 ladder model
is additionally  supported by  the fact that
the special point  $J_4^{c1}$ perfectly maps on the RS-HL phase boundary 
in the phase diagram of the frustrated spin-1/2 ladder model.\cite{weihong}  
Indeed, taking the parameters
$y_1=J_{\perp}^{'}/J_3^{'}$ and $y_2=J_4^{'}/J_3^{'}$ used in
Ref.~\onlinecite{weihong}, the established relations $J_s^{'}=J_s$ and
$J_a^{'}=-2\sqrt{2/3}\, J_a$  between the parameters of the 
original and the projected Hamiltonians take  the form
\begin{equation}\label{b}
y_1=\frac{J_{\perp}/J_3}{b_2J_4/J_3-b_1},\hspace{1cm}
y_2=\frac{b_2-b_1J_4/J_3}{b_2J_4/J_3-b_1},
\end{equation}
where $b_1=\sqrt{2/3}-1/2$ and $b_2=\sqrt{2/3}+1/2$.
Note that the change of $J_4$ (at fixed $J_{\perp}=J_3=1$) corresponds to
a run  in  the $(y_1,y_2)$ plane on the $ab$ line ( see Fig.~\ref{map}) defined
by $y_2=(b_1/b_2+1)y_1-b_1/b_2$. Following Ref.~\onlinecite{weihong},
we may identify  the position of the quantum phase transition with the  point
$J_4=J_4^{c1}\equiv 0.710$  where the spin-1/2 rung correlations
change their sign (see Fig.~\ref{c1}). We find that the $(y_1,y_2)$ image $A$ of the 
transition point  $J_4^{c1}$  maps perfectly on the phase boundary in
the  $(y_1,y_2)$ plane. In Figure~\ref{map}, we also show the symmetric point
$A^{'}$ obtained by  the coordinate transformations 
$y_1\rightarrow y_1/y_2$ and $y_2 \rightarrow 1/y_2$, which are related to the exchange 
symmetry $J_3\longleftrightarrow J_4$ of the Hamiltonian.  As expected, 
the symmetric point $A^{'}$ also lies on the phase boundary.
\begin{figure}[hbt]
\begin{center}
\includegraphics[width=8cm]{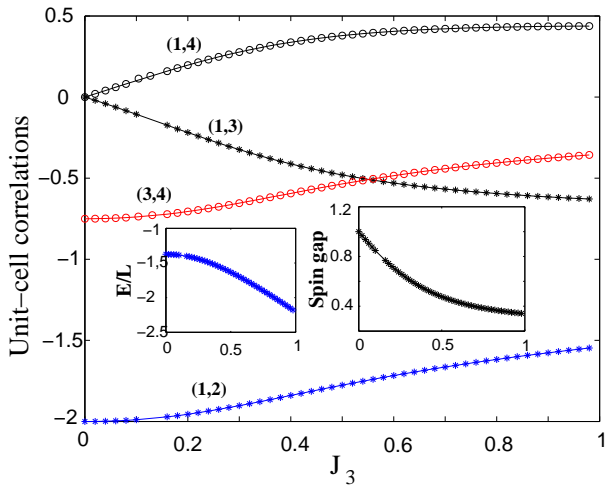}
\caption{\label{c2}
Unit-cell isotropic spin-spin  correlations of the model (\ref{h1}) 
as a function of $J_3$ ($J_4=0$, $L=100$).
The notations are defined in Fig.~\ref{c1}. 
The inset on the left shows the   ground-state energy per rung
\textit{vs.} $J_3$ and
the inset on the right shows the variation of the singlet-triplet
excitation gap with $J_3$.
}
\end{center}
\end{figure}

\subsection{Rung-singlet and Haldane-like phases}
\subsubsection{Rung-singlet phase}
The RS phase, originally studied in the two-leg spin-1/2 ladder
without  diagonal bonds,\cite{dagotto,barnes} 
is a non-degenerate singlet state with a finite singlet-triplet gap. The existence of a 
spin gap in this model can be easily anticipated by using a strong-coupling
analysis:\cite{barnes} For $J_3^{'}/J_{\perp}^{'}\ll 1$, the
ground state is a simple product of rung singlet bonds. The lowest rung excited
states are local triplets with a characteristic gap $\propto J_{\perp}^{'}$
which survives the perturbation in $J_3^{'}/J_{\perp}^{'}$. On the other
hand, the  perturbation produces an energy band (with a bandwidth $\propto J_3^{'}$) 
of triplet excitations.

The same physics  can be easily extracted from a strong-coupling
analysis of the mixed-spin ladder (\ref{h}). Instead of doing this, we present in
Fig.~\ref{c2} DMRG results for the short-range correlations  as a function of
$J_3$ ($J_4=0$).   
The state at $(J_3,J_4)=(1,0)$ is known to be gapped.\cite{trumperandgazza} 
The essential information in Fig.~\ref{c2} is that the curves are devoid of any features 
that might suggest a change of the phase. Thus we can assert that the phase at $J_3=1$ is smoothly 
connected to the phase at $J_3=0$,  which is a RS phase. The variation of the gap with $J_4$ is 
shown in Fig.~\ref{gaps}. We see  that the  gap  goes to zero around 
the point $J_4=0.710$ identified above as a phase transition point to another
singlet phase. Below we discuss in more detail the structure of the low-lying excitations
close to $J_4=J_4^{c1}$. 

\subsubsection{Haldane-like phase}
The discussed mapping of Eq.~(\ref{h}) on the frustrated spin-1/2 ladder model
suggests that the HL phase should  occupy  some region in the phase diagram
for $J_4 >J_4^{c1}$. To reveal the peculiarities of the suggested HL
phase -- as compared to the well-known Haldane phase of the periodic spin-1 Heisenberg chain
 -- notice that in the sector $[1,1,\ldots , 1]$
the Haldane state $|\Psi_H\rangle$  is the  exact 
ground state of the mixed-spin Hamiltonian (\ref{h}) at the symmetric point $J_3=J_4$.      
In the general case  ($J_3\neq J_4$), the energy of this state 
$E_H=\langle \Psi_H|{\cal H}|\Psi_H\rangle$ reads as
\begin{equation}\label{eh}
\frac{E_H}{L}=-\frac{J_1}{2}+\frac{J_2}{8} +\frac{1}{2}\left(
J_3+J_4\right)\varepsilon_H\, , 
\end{equation}
where $\varepsilon_H=-1.40148403897(4)$ is the the ground-state energy per
bond  of the periodic spin-1 Heisenberg chain.\cite{white3} Here, we have used the 
fact that the operator $V$ [Eq.~(\ref{v})] does not have   non-zero matrix 
elements in the sector $[1,1,\ldots , 1]$: In particular, we have
$\langle \Psi_H|V|\Psi_H\rangle =0$. The energy of the Haldane state  $E_H$
as a function of $J_4$ ($J_1=J_2=J_3=1$) is shown in Fig.~\ref{gse} (the $ab$
line). Interestingly, at the special point $J_4=J_4^{c2}\equiv 0.875$ -- also related  
to an abrupt change of the spin-1 rung correlations -- the DMRG estimate for the
ground-state energy of the Hamiltonian (\ref{h}) $E/L=-1.6899$  almost coincides with the
energy of the  Haldane state ($E_H/L=-1.6889$) obtained from Eq.~(\ref{eh}). 
As already mentioned above, the  numerical analysis implies that the special  
point $J_4^{c2}$ is a quantum phase-transition point from a singlet non-degenerate state 
to a state exhibiting a net magnetic moment. The above remarks suggest  that
the HL phase appears as  a  good candidate for the phase diagram of the mixed-spin
model. 

Further  qualitative information about the characteristics of this  phase 
can be extracted from a perturbative analysis starting from the symmetric
point $J_3=J_4$ and based on the  Haldane state in a periodic spin-1 chain. 
Note that in some interval ($J_4 < J_4^{c2}$) the parameter $J_a$, which
controls  the $V$ term in Eq.~(\ref{h}),  may be used as a small parameter 
(e.g., $J_a=0.0625$ for
$J_4=0.875$). Thus, up  to second order in $J_a$, the ground-state energy
takes the form $E=E_H-const\left(1-J_4\right)^2L$, where $const$ is some positive number
of order one. Qualitatively, this  result reproduces the  behavior of the
ground-state energy  in the interval $J_4^{c1}<J_4<J_4^{c2}$ extracted from the DMRG analysis 
(see Fig.~\ref{gse}). To some extent, this result  also validates the choice of 
$|\Psi_H\rangle$ as a starting unperturbed state.    

As compared to the Haldane state, some peculiarities of the HL phase can be 
revealed by looking at the first-order correction in $J_a$ to  the wave function 
$|\Psi_H\rangle$,
\begin{equation}\label{psi}
|\Psi\rangle=|\Psi_H\rangle+J_a\sum_{n\neq
0}\frac{\langle\Psi_n|V|\Psi_H\rangle}{E_0-E_n}|\Psi_n\rangle+{\cal O}\left(
J_a^2\right).
\end{equation}
Here the sum runs over the excited eigenstates
$|\Psi_n\rangle$  of the Hamiltonian (\ref{h}) at $J_3=J_4$, and $E_0\equiv
E_H$. The matrix elements of $V$ (see the Appendix) admit only two types of excited states
$(|\Psi_{1,2}\rangle)$ defined, respectively, in the sectors $[1,\ldots,1,0,0,1,\ldots,1]$ 
(two neighboring rungs in singlet states) an $[1,1,\ldots,1,2,0,1,\ldots,1]$ (one rung in a
 a quintet state an a neighboring rung in a singlet state). The weights of both
types of  defect configurations in the HL state change in the interval $J_4^{c1}<J_4
< J_4^{c2}$: While the  weight of the $|\Psi_{1}\rangle$ configurations
grows in a region around the transition point  $J_4^{c1}$, the $|\Psi_{2}\rangle$
configurations (containing spin-2 defects) become visible in the DMRG
result for the spin-1 rung correlations  only in a short interval preceding the 
transition to a magnetic state (see Fig.~\ref{c1}). Note that the observed
increase of the weight of the 
$|\Psi_{2}\rangle$ configurations formally contradicts the perturbation result in
Eq.~(\ref{psi}), which  predicts the opposite behavior. A reasonable
resolution for this is provided by the guess that close to the
transition point $J_4^{c2}$  some of the  eigenenergies $E_n$ 
related to the sector $[1,1,\ldots,1,2,0,1\ldots,1]$
soften. As of now we do not have firm  numerical results in favor of such a suggestion, 
although some  preliminary DMRG results, using open boundary conditions, seem to 
predict strong reductions of the singlet-quintet and triplet-quintet gaps close to
$J_4^{c2}$.
\subsubsection{The RS-HL transition}
\begin{figure}[hbt]
\begin{center}
\includegraphics[width=8cm]{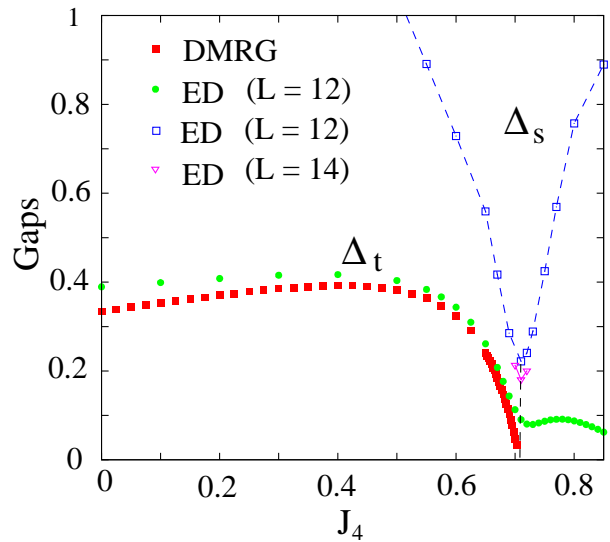}
\caption{\label{gaps}
DMRG and ED numerical results for the singlet-singlet ($\Delta_s$) and singlet-triplet
($\Delta_s$) excitation gaps in the mixed-spin model (\ref{h1}) \textit{vs.}
$J_4$. The DMRG data points correspond to extrapolated values of $\Delta_s$
obtained by a polynomial fit (up to $L=90$) for open boundary
conditions. The ED data concerns  periodic clusters with $L=10,12, 14$. 
}
\end{center}
\end{figure}

Turning to the region around the  transition point $J_4^{c1}$, it is
instructive to comment on our numerical results for the excitation gaps (Fig.~\ref{gaps})  in
the light of the discussed mapping  to the spin-1/2 ladder model. For the latter
model, it has been numerically established\cite{weihong} that (i) the lowest  
state above the singlet ground states close to the phase boundary is a singlet excitation  
and (ii) the low-lying triplet excitations are gapped in the whole region of the
phase diagram in Fig.~\ref{map}, including  the phase-transition
boundary. Such a structure of the low-lying excitations
is consistent with the established first-order quantum phase transition,
which is  described as a
level crossing of two  singlet ground states. As already mentioned, the
character of the RS-HL transition in the weak-coupling limit
($J_{\perp}^{'},J_{4}^{'}\ll J_{3}^{'}$) is still 
under debate.\cite{starykh,kim,liu} As a matter of fact, there are some
indications for a second-order RS-HL transition\cite{starykh} and
an intermediate dimer phase\cite{vekua,liu}, but the debate concerns only
the weak-coupling part of the phase boundary. Looking at the coordinates of the $A$
and $A^{'}$ images of the transition point $J_4^{c1}$ (Fig.~\ref{map}), it is clearly seen
that the discussed  RS-HL transition at $J_4=J_4^{c1}$ does not belong to the
weak-coupling region. Hence,  one may expect a first-order
RS-HL transition at  $J_4^{c1}$ related to a level crossing of singlet
ground states.   

Figure~\ref{gaps} presents our  numerical (DMRG and ED) results for the singlet
($\Delta_s$) and triplet ($\Delta_t$) gaps of the lowest excited modes above 
both singlet ground states. Let us first discuss the ED data for the gaps.  
As clearly seen, both  minima, related to the $\Delta_s$
and $\Delta_t$ data points,  are located close to  the expected transition point at
$J_4=0.710$. More importantly, an extrapolation of the ED data  for $J_4=
0.710$ implies 
that the $\Delta_s$ points  scale  to smaller values than $\Delta_t$. This
observation is consistent with the expected low-energy  structure close the  
first-order transition point between the RS and HL phases. 

Turning to the DMRG results for $\Delta_t (J_4)$, one  observes that the triplet gap
of the  RS phase takes very small values close to the suggested transition point
($J_4=0.710$). We could not conclusively exclude the possibility of
a gapless triplet excitation  at the transition point. In any case, such a
behavior indicates some peculiarities of the RS-HL transition in  the mixed-spin
system,  as compared to the uniform-spin case. Another issue to be noticed
is  the  steep (but definitely finite) slope of the function $\Delta_t (J_4)$ 
at the transition point. This  suggests a relatively large correlation
length of this triplet excitation close to $J_4^{c1}$.
\subsection{Ferrimagnetic phases}
\begin{figure}[t]
\begin{center}
\epsfig{figure=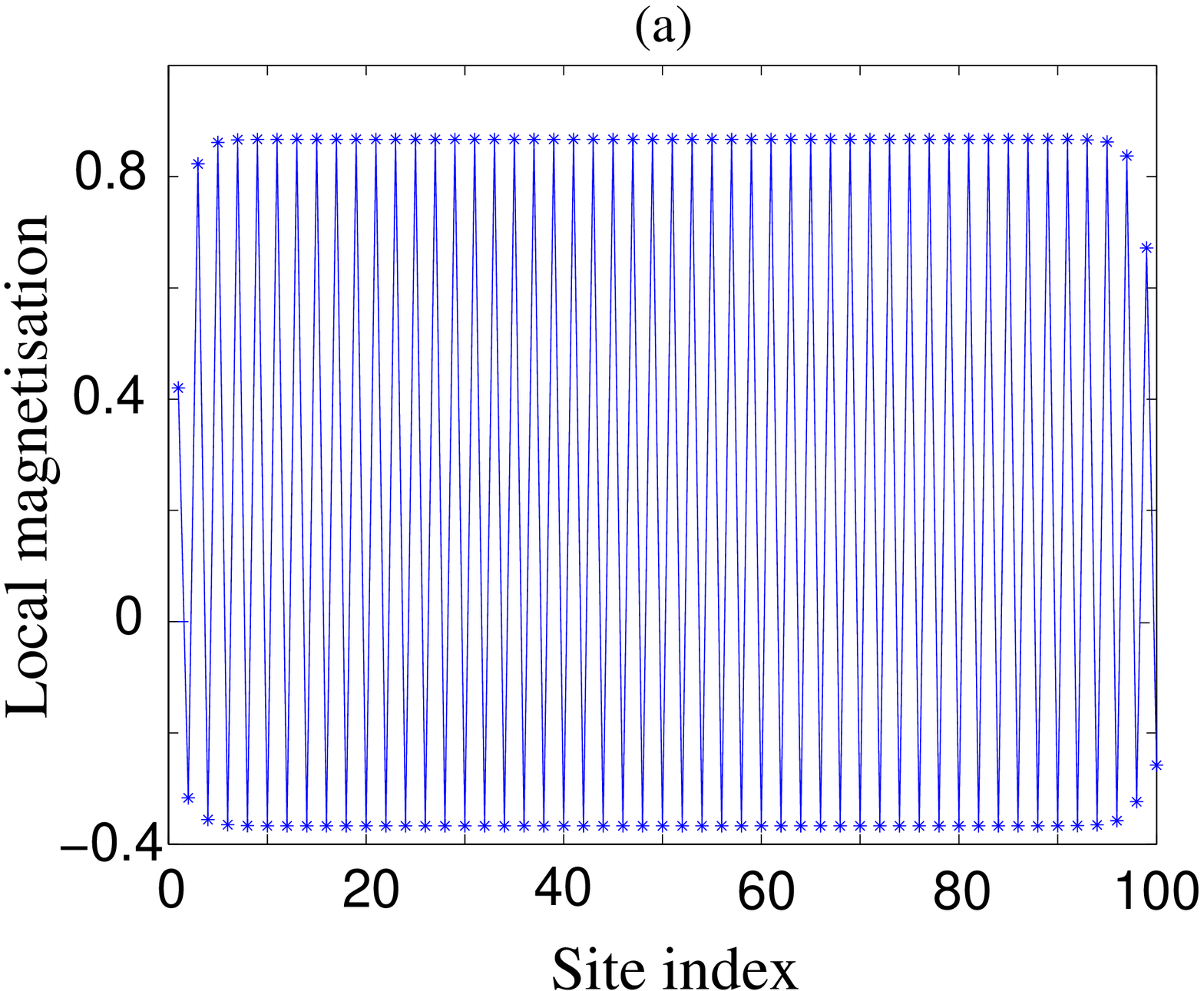,width=4cm}
\epsfig{figure=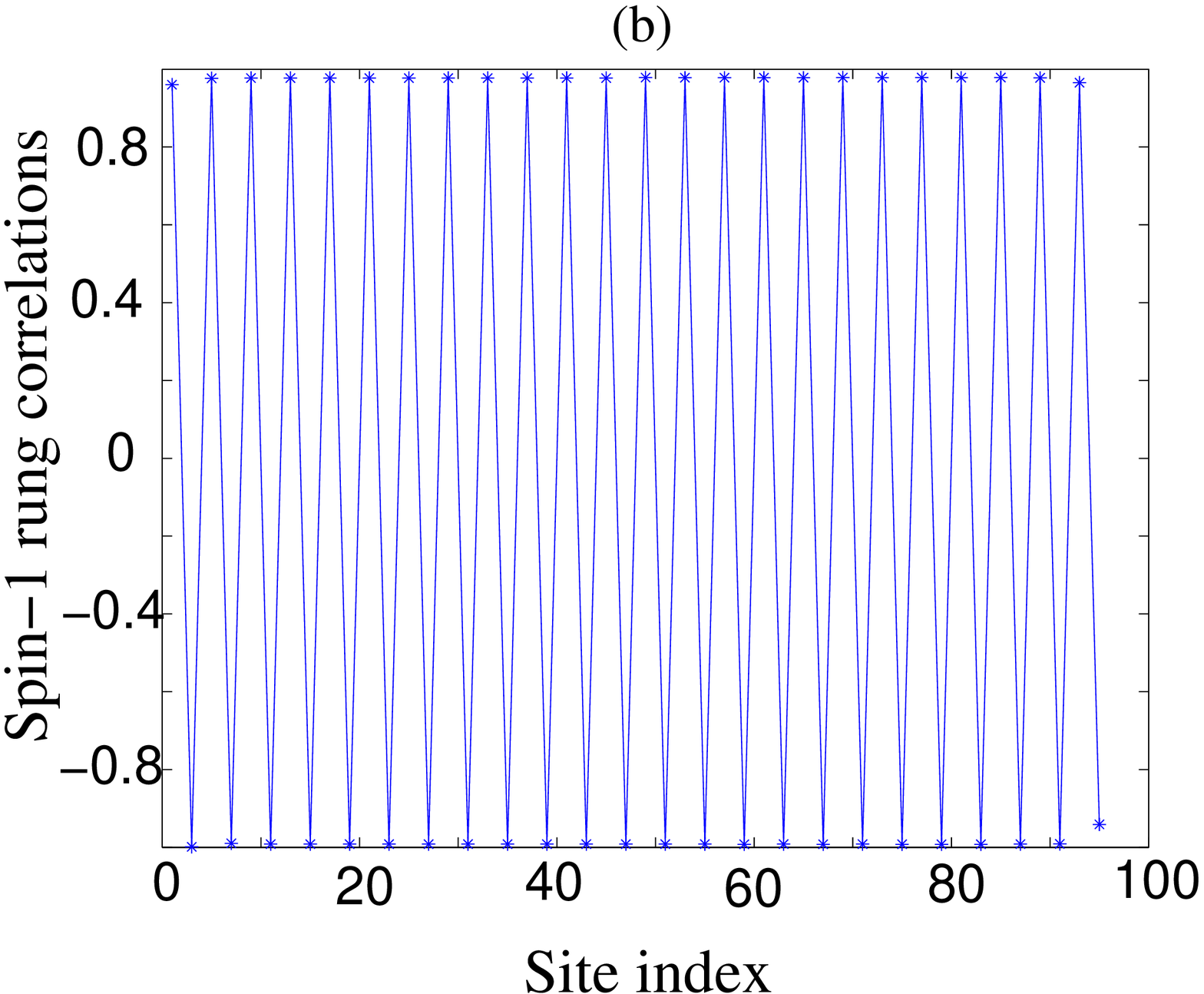,width=4cm}
\caption{\label{ferri} (a) The local magnetizations $\langle
s_{1,2n}^z\rangle$ and $\langle\sigma_{1,2n+1}^z\rangle$ ($n=1,\ldots,50$) 
along the first leg as a function of the site  index.
The data shown is for $J_4=1.55$.
(b) The spin-1 rung correlations along the length of the 
ladder ($L=100$)  at $J_4=0.90$.
The values show a clear alternation between $\approx 1$ and $\approx -1$
which indicates a two sublattice structure and a doubled unit cell containing
four rungs.
}
\end{center}
\end{figure}

Looking at the DMRG results for the short-range correlations (Fig.~\ref{c1}), it is easy to
realize that  a ferrimagnetic phase, closely related  to the ferrimagnetic ground state of an 
antiferromagnetic Heisenberg chain with alternating $(2,1)$ spins, is stabilized 
around the symmetric point $J_4=1$. Exactly at $J_4=1$, the ground state of
the Hamiltonian  (\ref{h}) belongs to the sector  $[1,2,\ldots,1,2]$, so that
both models are equivalent  in the low-energy sector of the spectrum. The
discussed ferrimagnetic phase ($F_1$) exhibits the magnetic 
moment per rung $M_0=1/2$ and survives almost in the entire region after
$J_4^{c2}$, excluding some narrow interval in the vicinity of the latter
point.   
This is also seen in Fig.~\ref{ferri}(a) which shows a typical  behavior of
the local magnetizations $\langle s_{1,2n}^z\rangle$ and
$\langle\sigma_{1,2n+1}^z\rangle$ ($n=1,\ldots,L/2$)  along the first leg at $J_4=1.55$. 
The values of the spin-1 and spin-1/2 magnetic moments are $0.866950$  and
$-0.366950$,
respectively. We see that the sum of the local magnetic moments is $1/2$, as expected in a 
Lieb-Mattis type ferrimagnetic state with a quantized magnetic moment per rung $M_0=1/2$.
The deviations at the end are essentially because of open boundary conditions.
We have verified numerically that these values do not change much after $J_4=1$.

For the region close to $J_4^{c2}$, the  DMRG results presented in
Fig.~\ref{ferri}(b) demonstrate the appearance of another ferrimagnetic phase
($F_2$)  in  a narrow range of $J_4$ starting from the transition point $J_4^{c2}=0.875$ 
and terminating at $J_4^{c3}=0.975$. The $F_2$ phase is characterized by the
magnetic moment per rung $M_0=1/4$. As clearly seen  in Fig.~\ref{ferri}(b), in
the $F_2$ phase the space variation of the spin-1 rung correlations
follow strictly the periodicity of the spin structure in the sector $[2, 1, 1, 1,\ldots, 2,1,1,1]$.  
Such a  breaking of the translational symmetry is also seen in  the
inset of Fig.~\ref{c1}, where on the vertical axis we have plotted the magnitude of the difference of 
the spin-1 rung correlations in two neighboring unit cells for all values of $J_4$. 
Clearly,  the $F_2$ phase represents a  two-fold degenerate ground state,
which is invariant under the translation by two lattice periods. Our numerical analysis does not
support the appearance of  ferrimagnetic phases with larger
periods.          
\section{Conclusion}\label{sum}
In conclusion, we have analyzed the combined effect of the quantum fluctuations
and the competing interactions in  a mixed-spin ladder composed of spin-1
and spin-1/2 rungs which is closely related to a recently synthesized quasi-1D
ferrimagnetic material. A comparison of the classical and quantum phase diagrams
reveals the following changes in the related quantum system.
As expected, the classical ferrimagnetic phase also presents in the quantum phase
diagram, but there appears another two-fold degenerate ferrimagnetic state which breaks 
the translational symmetry. As may be expected, the classical N\'eel state
does not survive quantum fluctuations. More interestingly, the classical canted state also 
completely disappears. This is in contrast to some other 1D spin systems
exhibiting classical canted states,\cite{ivanov_09}  where this type of classical magnetic 
order  partially survives quantum fluctuations. In the present case, 
both the classical long-range ordered  states are replaced by two
singlet non-degenerate gapped states (RS and HL).             

Turning to the weakly frustrated region, it has been established that the behavior of the
system strongly resembles that  of a two-leg spin-1/2 Heisenberg ladder with
frustrating diagonal interactions. However, concerning the quantum phase transition
between the RS and HL phases, we have found  a few indications
demonstrating some peculiarities (such as  the extremely small triplet gap at
the transition point) of the mixed-spin system. These issues deserve 
further investigations. 

Finally, although the available experimental 
results on the ferrimagnetic ladder material 
Fe$^{II}$Fe$^{III}$ (trans-1,4-cyclohexanedicarboxylate)
seam to point toward the realization of the  $F_1$ ferrimagnetic state,\cite{zheng} 
a detailed comparison with the experiment  requires  a more extensive analysis  
of the quantum phase diagram including, e.g.,  different rung couplings $J_1\neq J_2$,
different pairs of rung spin magnitudes,  and some anisotropies. Concerning
the condition $J_3=1$, as shown in Fig.~\ref{map} it simply restricts the
path in the more general parameter  space  ($J_3\neq 1$) to a straight line
crossing one and the same phase boundary. Therefore, there should be  a
relatively  large region  with $J_3\neq 1$ showing the same structure of the phase diagram. 
As to the second  restriction ($J_1=J_2$), its removal may be generally expected to bring new
quantum spin phases. However, in both cases we have numerically checked that
relatively small deviations from the conditions  $J_1=J_2=J_3$ do not bring
qualitative changes on the established quantum phase diagram.

\acknowledgments{
This work has been supported by the Bulgarian Science Foundation (Grant DO02-264/18.12.08).
J. R. is also indebted to the DFG for financial support (project
RI615/16-1).
V. R. C. thanks the MPIPKS in Dresden (Germany) for financial support and computational 
resources for most of the duration of the project and acknowledges being supported
in part at the Technion by a Fine Tust when the manuscript was being finalized.
He thanks Andreas L\"auchli and Masaaki Nakamura for useful discussions.
}
\appendix*
\section{Projection onto the spin-1/2 ladder}
We have to project the spin-1 rung states onto the states of the spin-1/2
rungs. To this end, we use the projection operator $P=P_1P_2,\ldots,P_L$,
where the rung projection operator $P_n$ reads as
\begin{equation}
P_n=\sum_\alpha |T^{\alpha}_{2n}\rangle\langle T^{\alpha}_{2n}|,\,\,
\alpha =0,x,y,z.
\end{equation}
Here $|T^{0}_{2n}\rangle$ denotes the singlet state of the $2n$th 
spin-1 rung and $|T^{k}_{2n}\rangle =(i/\sqrt{2})\epsilon^{klm}|l\rangle |m\rangle$ 
are the triplet states of the same rung in a vector basis which is a tensor
product of the vector bases of the spin-1 objects (i.e., $|x \rangle$,
$|y \rangle$, and $|z \rangle$). In the following, the Greek indices take the values $0,x,y$, and
$z$, whereas the Latin ones --  $x,y$, and $z$.  

Up to first order in $J_a$, the projected Hamiltonian reads as
\begin{equation}
{\cal H}_{eff}=P{\cal H}P.
\end{equation}
By using the expressions for the matrix elements $\langle T^{m}_{2n}
|{\bs s}_{2n}^2|T^{n}_{2n}\rangle =2\delta^{m n}$, 
$\langle T^{0}_{2n}|{\bs L}_{2n}^k|T^{0}_{2n}\rangle = 
\langle T^{m}_{2n}|{\bs L}_{2n}^k|T^{l}_{2n}\rangle =0$, and 
$\langle T^{m}_{2n}
|{\bs L}_{2n}^k|T^{0}_{2n}\rangle =-2\sqrt{2/3}\delta^{m k}$, one obtains 
\begin{equation}
P_n{\bs s}_{2n}^2 P_n=2\sum_k |T^{k}_{2n}\rangle\langle
T^{k}_{2n}|={\bs \sigma}_{2n}^2,
\end{equation}
where ${\bs \sigma}_{2n}$ is an effective  rung-1/2 spin operator,
and
\begin{eqnarray}
&&P_nV_nP_n=\nonumber \\
\hspace{-1cm} 
&&-2\sqrt{\frac{2}{3}}\sum_k 
\left[ |T^{0}_{2n}\rangle\langle T^{k}_{2n}|+
|T^{k}_{2n}\rangle\langle T^{0}_{2n}|\right]
\left( l^k_{2n-1}+l^k_{2n+1}\right).\nonumber
\end{eqnarray}

Note that the operator in the square brackets is an effective 
${\bs l}_{2n}$ rung vector operator for spin-1/2 rungs.
Summing the above results, we obtain the  effective spin-1/2 
ladder model presented in Eq.~(\ref{he}).


\end{document}